\documentclass[a4paper,11pt]{article}
\usepackage{setspace}
\usepackage{feynmf}
\usepackage{amsmath,amssymb,amsthm}
\usepackage[totalwidth=17cm,totalheight=24cm]{geometry}
\usepackage{empheq}
\usepackage{graphicx}
\newcommand{\ket}[2]{ {\langle #1\,#2\rangle} }
\newcommand{\bra}[2]{ {[#1\,#2]}}

\newcommand{\im}{{\mathrm i}}

\newcommand{\C}{{\mathbb C}}

\newcommand{\cK}{{\mathcal K}}
\newcommand{\cI}{{\mathcal I}}
\newcommand{\cJ}{{\mathcal J}}
\newcommand{\cT}{{\mathcal T}}

\newcommand{\Res}{{\rm Res}}

\newcommand{\be}{\begin{eqnarray}}
\newcommand{\ee}{\end{eqnarray}}

\begin{document}

\title{Pure connection formalism for gravity: Recursion relations}
\author{Gianluca Delfino, Kirill Krasnov and Carlos Scarinci \\ \\ \it{School of Mathematical Sciences, University of Nottingham}\\ \it{University Park, Nottingham, NG7 2RD, UK} }
\date{October 2014}
\maketitle
\begin{abstract}\noindent
In the gauge-theoretic formulation of gravity the cubic vertex becomes simple enough for some graviton scattering amplitudes to be computed using Berends-Giele-type recursion relations. We present such a computation for the current with all same helicity on-shell gravitons. Once the recursion relation is set up and low graviton number cases are worked out, a natural guess for the solution in terms of a sum over trees presents itself readily. The solution can also be described either in terms of the half-soft function familiar from the 1998 paper by Bern, Dixon, Perelstein and Rozowsky or as a matrix determinant similar to one used by Hodges for MHV graviton amplitudes. This solution also immediate suggests the correct guess for the MHV graviton amplitude formula, as is contained in the already mentioned 1998 paper. We also obtain the recursion relation for the off-shell current with all but one same helicity gravitons. 
\end{abstract}

\section{Introduction}

In \cite{Krasnov:2011pp}, \cite{Krasnov:2011up} $\Lambda\not=0$ General Relativity and, more generally, a certain infinite-parametric class of parity-odd theories of interacting massless spin two particles, were described as diffeomorphism invariant ${\rm SO}(3)$ gauge theories. Papers \cite{Delfino:2012zy}, \cite{Delfino:2012aj} gave the rules for computing (Minkowski space) graviton scattering amplitudes using this formalism; the simplest such amplitudes (graviton-graviton scattering) were computed in \cite{Delfino:2012aj}. 

In this paper we use this gauge-theoretic formalism for gravity to compute more general graviton amplitudes. Thus, we use Berends-Giele-type \cite{Berends:1987me} recursion relations for the off-shell currents to investigate and compute the current with all on-shell gravitons of the same helicity. This current is arguably the simplest object one has in any massless theory. Indeed, it is the computation of this current that led the authors of \cite{Berends:1987me} to the proof of the Parke-Taylor formula \cite{Parke:1986gb} for MHV gluon amplitudes. Similarly in gravity, the fact that all on-shell legs are of the same helicity leads to many simplifications. The resulting recursion relation can then be solved in full generality. The answer contains all the essential features of the MHV graviton scattering amplitude, as is known already from the work \cite{Bern:1998sv}. In fact, once this current is computed, it suggests a similar-type formula for the MHV graviton scattering amplitude, as was first described in \cite{Bern:1998sv}. 

The MHV amplitudes formula in \cite{Bern:1998sv}, which is arguably the most beautiful available formula for these amplitudes, was rediscovered in \cite{Nguyen:2009jk}. Later, a representation in terms of a matrix determinant was found in \cite{Hodges:2012ym}. A simple application \cite{Feng:2012sy} of the matrix tree theorem relates the sum over trees representation \cite{Bern:1998sv}, \cite{Nguyen:2009jk} with the determinant representation of Hodges \cite{Hodges:2012ym}. 

Our computation of the off-shell gauge-theoretic current leads to an essentially the same recursion relation that appears in \cite{Bern:1998sv}. An application of the matrix tree theorem also allows to represent the current as a certain matrix determinant. As is explained in \cite{Krasnov:2013wsa}, and will be reviewed below, there is also a BCFW-type recursion relation that can be written for such quantities. 

We also investigate the Berends-Giele-type recursion relation for the current with all but one on-shell leg of the same helicity. This is the current that could be used to compute the MHV amplitude directly, by setting the off-shell leg on-shell. We manage to transform the arising recursion relation to a reasonably simple form. However, the low graviton number cases do not appear to suggest any form of the general answer. Thus, we were unable to find the general solution. 

The computation of the all same helicity on-shell legs current presented in this paper can be viewed as the simplest available derivation of the MHV amplitude formula using Feynman diagrams. Indeed, the input of the computation presented in \cite{Bern:1998sv} is the formula for the cubic graviton vertex as a product of two gauge-theory vertices. This requires quite non-trivial manipulations with the Einstein-Hilbert action to establish (introduction of auxiliary field etc.). In contrast, in our formalism the simple cubic vertex responsible for the MHV amplitude comes out directly by expanding the action, without too much thought. Thus, the derivation presented in this paper is more self-contained than the one in \cite{Bern:1998sv}. 

We also believe that the derivation presented in this paper is valuable and timely for the following reason. It is often stated that Feynman diagrams carry a lot of unnecessary information, and only the modern on-shell methods can efficiently compute amplitudes. There is certainly truth to these statements. It is worth stressing however that it is the Feynman diagram derived Berends-Giele-type recursion relation that naturally and simply leads to the most beautiful sum over trees \cite{Bern:1998sv}, \cite{Nguyen:2009jk} representation for the MHV amplitudes. Certainly, once such a representation is found one can prove it using the powerful tool of BCFW recursion relations \cite{Britto:2005fq}, as was done in \cite{Nguyen:2009jk}. However, it is very hard to see that the solution to BCFW recursion organises itself into a sum over trees, while this directly follows from the Berends-Giele Feynman diagram derived recursion. Thus, the main message that we would like a reader to take from this paper is that Feynman diagrams, if used correctly, can present structures invisible by the pure on-shell methods. 

The paper is organised as follows. We start by reviewing the gauge-theoretic description of gravity Feynman rules. These are derived in \cite{Delfino:2012zy}, \cite{Delfino:2012aj}, but will be presented here in a self-contained fashion, in the amount needed for our computations. We set up the Berends-Giele-type recursion for the all same helicity on-shell legs current in Section \ref{sec:all-minus}. Its solution readily presents itself by working out the few first amplitudes. We don't need to give a proof of the general formula here, because it is already spelled out in \cite{Krasnov:2013wsa} in greater generality. In Section \ref{sec:one-plus} we derive and analyse the recursion relation for the all but one same helicity on-shell legs current. We conclude with a brief discussion.

\section{Feynman rules}

We take the gauge-theoretic Feynman rules from \cite{Delfino:2012aj}, but describe them here in a slightly more formal, and convenient for computations language. 

In the gauge-theoretic approach to gravity one uses a certain parity asymmetric representation of the Lorentz group to describe gravitons. We remind the reader that the finite-dimensional irreducible representations of the Lorentz group are of the type $S_+^k\otimes S_-^n$, where $S_\pm$ are the fundamental (complex conjugate to each other in the case of Minkowski signature) two-dimensional spinor representations. The representations $S_+^k, S_-^n$ are the $k$-th and $n$-th symmetric tensor products of the corresponding fundamental representations. Thus, ${\rm dim}(S_\pm^k)=k+1$. 

In the usual metric description one uses the representation $S_+^2\otimes S_-^2$ to describe gravitons. Indeed, the trace free part of the metric perturbation $h_{\mu\nu}$, once the space-time indices are translated into the spinor ones, lives precisely in $S_+^2\otimes S_-^2$. Note that this space is parity-symmetric, and thus contains (in the case of Minkowski signature) real elements. These correspond to the trace free parts of perturbations of the real metrics. 

In contrast, in the gauge-theoretic formalism one uses the space $S_+^3\otimes S_-$, or the parity related space $S_+\otimes S_-^3$ if one prefers. Note that this space is not parity-symmetric, and the operation of complex conjugation sends elements of  $S_+^3\otimes S_-$ to elements of $S_+\otimes S_-^3$. Thus, there are no real elements in $S_+^3\otimes S_-$, and the reality conditions are more non-trivial to impose. As is explained in details in \cite{Delfino:2012zy}, the reality conditions are given by a certain second-order differential equation on $a\in S_+^3\otimes S_-$, which is in a certain sense an analog of the usual first order Dirac equation. Alternatively, one can view the reality condition as the statement that the metric, constructed as a derivative of the connection $a$, is real. The reality conditions involving derivatives bring with themselves a mass scale. This mass scale turns out to be related to the radius of curvature of the background. Thus, this formalism for describing gravitons only works on non-zero scalar curvature backgrounds. This brings in a complication that Minkowski space graviton scattering amplitudes have to be computed by a somewhat schizophrenic procedure where one does all computations in Minkowski space and uses the usual Fourier transform, but at the same time remembers that some quantities came from a constant curvature space, and so keeps factors of the cosmological constant in the intermediate formulas. These rules were derived in  \cite{Delfino:2012aj}, but once adopted, are in no way more involved than the familiar Feynman rules of e.g. Yang-Mills theory. 

Thus, the gravitons in our formalism are described by a field $a_{ABCA'}\in S_+^3\otimes S_-$, here $A,B,C,\ldots$ are the unprimed spinor indices used to denote elements of $S_+$, and $A',B',C',\ldots$ are the primed indices used to label spinors from $S_-$. The propagator is the usual scalar propagator times the identity matrix in the space $S_+^3\otimes S_-$:
\be
{\rm propagator} \,\, = \frac{1}{\im k^2} \, {\rm Id}_{S_+^3\otimes S_-}.
\ee

We then need to describe the two polarisation spinors. These are given just by what could be guessed from similar expressions in the metric formalism:
\be\label{polar}
\epsilon^-_{ABCA'} = M \, \frac{q_A q_B q_C k_{A'}}{\ket{q}{k}^3}, \quad \epsilon^+_{ABCA'} = \frac{1}{M} \, \frac{k_A k_B k_C p_{A'}}{\bra{p}{k}},
\ee
where 
\be
\ket{k_1}{k_2} := k_1^A k_{2 A}, \qquad \bra{k_1}{k_2} := k_{1 A'} k_2^{A'}
\ee
are the spinor contractions. Note that the polarisation spinors (\ref{polar}) are homogeneous degree zero in the reference spinors $q_A, p_{A'}$ as they should be. Note also that the factors of $M$ in the negative helicity and $1/M$ in the positive helicity are required for dimensional reasons: the helicity spinors (\ref{polar}) should be dimensionless. Finally, the spinors $k_A, k_{A'}$ are the momentum spinors $k_{AA'}=k_A k_{A'}$ appearing as the "square root" of the corresponding null momentum vector. 

We also need an expression for the cubic vertex. As is usual in such context, one selects all the reference spinors of the negative helicity gravitons to be the same spinor $q_A$. If all on-shell legs are those of negative helicity, the choice of $q^A$ remains arbitrary. If there is one positive helicity on-shell graviton, it is convenient to take $q^A$ to be equal to the momentum spinor of the positive helicity graviton. Then there are so many factors of the spinor $q_A$ that is very difficult to obtain a non-zero contraction. Very simple count shows that one needs the maximal possible number of derivatives to be present, which is the case when there are only cubic vertices. Furthermore, there are just two cubic vertices containing the maximum number of derivatives in the family of gravity theories of interest. It is easy to see that one of them cannot contribute, because it will necessarily make $q$'s contract. Thus, the only vertex that can contribute is as follows
\be\label{vertex}
{\rm vertex} = \frac{\kappa}{2M} ( \partial a, [\![ \delta a, \delta a]\!]).
\ee
We also note that this is the only maximal number of derivatives vertex that is present in the case of General Relativity. Thus, all our infinite-parameter family of gravity theories will share the same MHV graviton scattering amplitudes. This is clear even before any computation of these amplitudes is performed. The vertex (\ref{vertex})  is the relevant part of the Lagrangian with the factor of the imaginary unit in front ${\rm vertex} = \im {\cal L}$, and $\kappa^2= 32\pi G$. We have written it in terms of derivative operators, which have to be replaced by the factors of $\pm \im$ times the momentum flowing on the corresponding line of the diagram. The differential operators appearing here are defined as
\be
\partial: S_+^3 \otimes S_- \to S_+^4, \qquad (\partial a)^{ABCD}:= \partial^{(A}_{M'} a^{BCD)M'}\\
\nonumber
\delta: S_+^3\otimes S_- \to S_+^2 \otimes S_-^2, \qquad (\delta a)^{ABA'B'} := \partial^{E(A'} a_E{}^{ABB')}.
\ee
The product $[\![ , ]\!]$ of two elements of $S_+^2\otimes S_-^2$ is defined as
\be
[\![ , ]\!]: (S_+^2\otimes S_-^2) \otimes (S_+^2\otimes S_-^2) \to S_+^4, \qquad
[\![ h_1 , h_2]\!]^{ABCD} := h_1^{(AB}{}_{A'B'} h_2^{CD)A'B'}.
\ee
This product is commutative. We have denoted it by the double commutator symbol because a single contraction of a pair of spinor indices represents the commutator. Here we have a double contraction, hence the notation. Finally, the round brackets stand for the inner product in $S_+^4$:
\be
({},{}): S_+^4 \otimes S_+^4 \to \C, \qquad (\psi_1,\psi_2) := \psi_1^{ABCD} \psi_{2 ABCD}.
\ee
 
The only other convention that we need is the rule that the positive helicity gravitons are slightly massive:
\be\label{massive}
k^{AA'} = k^A k^{A'} + M^2 \frac{p^A p^{A'}}{\ket{p}{k}\bra{p}{k}}.
\ee
Here the reference spinors $p^{A'}, p^A$ are the same as used in the corresponding positive helicity spinor (\ref{polar}). It is this rule, together with the factors of $M$ appearing in all formulas, that remember about the fact that this perturbation theory was originally set up in de Sitter space of cosmological constant $\Lambda/3=M^2$. However, as well shall see, no factors of $M$ will remain in the final results for what can then be interpreted as the Minkowski space amplitudes. 

Let us now derive some simple consequences of the above rules. First, let us compute the action of the operators $\delta, \partial$ on the helicity states. Assuming that that particles are incoming so that the derivative operator gets replaced with a factor of $-\im k_{AA'}$ we get:
\be
(\partial \epsilon^-)^{ABCD}=0, \qquad (\partial \epsilon^+)^{ABCD} = \frac{\im}{M} k^A k^B k^C k^D, \\ \nonumber
(\delta \epsilon^-)^{ABA'B'} = \im M \frac{q^A q^B k^{A'} k^{B'}}{\ket{q}{k}^2}, \qquad
(\delta \epsilon^+)^{ABA'B'} = -\im M \frac{k^A k^B p^{A'} p^{B'}}{\bra{p}{k}^2}.
\ee
Note that the application of the $\delta$ gives the familiar metric helicity spinors. The application of $\partial$ gives the self-dual part of the Weyl curvature when applied to the positive helicity graviton, and zero when applied to negative helicity (because the negative helicity does not have the self-dual part of Weyl curvature). Note that to get the last relation we have used (\ref{massive}). 

\section{All minus current recursion relation}
\label{sec:all-minus}

The Berends-Giele recursion first appeared in paper \cite{Berends:1987me} in the context of Yang-Mills theory. The general recursion relation involves both the 3-valent and 4-valent YM vertices. For MHV amplitudes only the 3-valent vertices can contribute, which results in a particularly simple recursion. This recursion relation can then be used to prove the Parke-Taylor formula \cite{Parke:1986gb}. 

For gravity the Feynman rules of the usual metric formulation are too complicated to write any analog of the Berends-Giele recursion relation, even for the MHV amplitudes where only the 3-vertices contribute. See, however, \cite{Bern:1998sv}. Thus, in the case of gravity the general MHV graviton scattering amplitude is usually determined in a different way. Historically the first was the work \cite{Berends:1988zp}, that used the string-inspired gauge-gravity scattering amplitudes relation \cite{Kawai:1985xq} to determine the general n-graviton MHV amplitude. Another route to the general n-graviton MHV formula is via the BCFW recursion relations \cite{Britto:2005fq}. These were applied to graviton scattering in \cite{Bedford:2005yy} and \cite{Cachazo:2005ca}. Yet another self-contained derivation of the graviton MHV formula is given in \cite{Mason:2008jy} and uses the twistor construction. 

In this section we set up a Berends-Giele-type recursion relations for determining the MHV graviton amplitudes. In this paper we adopt the terminology that an MHV amplitude is one with at most two plus helicity gravitons, i.e. it is the mostly minus amplitude. From general considerations (counting the number of reference spinors $q^A$) it is easy to deduce that only the 3-valent vertices give a contribution to the MHV amplitudes. This is completely analogous to what happens in Yang-Mills theory and gravity in the usual metric formulation, where it is easy to show that only the diagrams with the maximal possible number of derivatives can contribute, and these are the 3-valent diagrams. It is the fact that only 3-valent vertices contribute, as well as the simplicity of our vertex (\ref{vertex}), that allows to write down a simple Berends-Giele-type recursion. 

\subsection{Glueing together two negative helicity gravitons}

Thus, we would like to analyse the current with all negative helicity on-shell legs. This is defined as an amplitude with $n$ on-shell negative helicity gravitons, and one off-shell leg. The convention is that the final propagator for the off-shell leg is included in the current, so that such current can be glued together directly, with no extra propagators to be added, via vertices. By definition, the current for a single graviton is the corresponding helicity spinor that we rewrite in a convenient for later fashion
\be\label{J-1}
J(1)_{ABCD'}\equiv\epsilon^-(1)_{ABCD'}=M q_A q_B q_C 1_{DD'} q^D A(1).
\ee
Here we have introduced a convenient convention $(k_i)_{AA'} \equiv i_{AA'}$, and denoted
$$A(1)=\frac{1}{\ket{1}{q}^4}.$$

Let us now consider the process of glueing two such currents (negative helicity states) in the vertex (\ref{vertex}). Because the application of $\partial$ to a negative state gives zero, the only way to get a non-zero result is to insert the two negative helicity states into the two symmetric legs of the vertex, i.e. those where it is the derivative operator $\delta$ that gets applied. Above we have already computed the result of action of $\delta$ on a negative helicity state. In the present notations this is rewritten as
\be\label{dJ1}
(\delta J(1))^{ABA'B'} = \im M q^A q^B  q^E q^F 1_E{}^{A'} 1_F{}^{B'} A(1).
\ee
The double commutator of two such quantities from $S_+^2 \otimes S_-^2$ gives the following element of $S_+^4$
\be
[\![\delta J(1) ,\delta J(2)]\!]^{ABCD} = (\im M)^2 q^A q^B q^C q^D ( q^E q^F 1_E{}^{A'} 1_F{}^{B'}) (q^P q^Q 2_{PA'} 2_{QB'} ) A(1) A(2).
\ee
Applying the final $\partial$ derivative that gives a factor of $\im (1+2)_{AA'}$, and appending the propagator $1/2\im \ket{1}{2}\bra{1}{2}$ at the end, gives the following result for the current for two gravitons
\be\label{J-12}
J(1,2)_{ABCD'} =  M \left(\frac{\kappa}{2}\right) q_A q_B q_C (1+2)_{DD'} q^D A(1,2),
\ee
where
\be
A(1,2) = \frac{1}{\ket{1}{2}\bra{1}{2}} (\ket{q}{1}\ket{q}{2}\bra{1}{2})^2 A(1) A(2).
\ee
Thus, the current for two gravitons has the same spinor structure as the single graviton state (\ref{J-1}).

\subsection{General recursion}

This result immediately generalises. It is easy to see that the only way to glue structures as in (\ref{J-12}) via the vertex (\ref{vertex}) is to insert these currents into the $\delta$-legs of the vertex. This shows that the general all minus current has the following structure
\be
J(\mathcal{K})_{ABCD'}=M\left(\frac{\kappa}{2}\right)^{|\mathcal{K}|-1}q_A q_B q_C (\sum_{\mathcal{K}}k)_{DD'} q^D A(\mathcal{K}),
\ee
where $\mathcal{K}$ is the set of negative helicity momenta and the higher $A$-coefficients are computed using the following recursion
\be\label{rec_rel_A}
A(\mathcal{K})
=\frac{1}{\sum_{i<j\in\mathcal{K}}\ket{i}{j}\bra{i}{j}}\sum_{\mathcal{I}\sqcup\mathcal{J}=\mathcal{K},|\mathcal{I}|\leq|\mathcal{J}|}A(\mathcal{I})A(\mathcal{J})\left(\sum_{a\in\mathcal{I},b\in\mathcal{J}}\ket{a}{q}\ket{b}{q}\bra{a}{b}\right)^2
\ee
The sum runs over disjoint sets of momenta $\mathcal{I}$ and $\mathcal{J}$ whose union is $\mathcal{K}$ in such a way as to not repeat configurations, which we denote by the symbol $|\mathcal{I}|\leq|\mathcal{J}|$.

An important fact about this recursion relation is that at every step the prefactor in front of the above formula, which is proportional to $1/(\sum_i k_i)^2$, can be cancelled out with an analogous factor appearing in the numerator. In other words, the numerator in (\ref{rec_rel_A}) is always proportional to $(\sum_i k_i)^2$. 

The general solution to the recursion (\ref{rec_rel_A}) is easily guessed once the first few cases are worked out:
\be\label{As}
A(1,2) = \frac{1}{\ket{1}{q}^2\ket{2}{q}^2} \frac{\bra{1}{2}}{\ket{1}{2}}, \\ \nonumber
A(1,2,3)=\frac{1}{\ket{1}{q}^2\ket{2}{q}^2} \frac{\bra{1}{3}\bra{3}{2}}{\ket{1}{3}\ket{3}{2}}+\frac{1}{\ket{1}{q}^2\ket{3}{q}^2} \frac{\bra{1}{2}\bra{2}{3}}{\ket{1}{2}\ket{2}{3}}+\frac{1}{\ket{2}{q}^2\ket{3}{q}^2} \frac{\bra{2}{1}\bra{1}{3}}{\ket{2}{1}\ket{1}{3}} .
\ee
There is clearly a simple pattern here, and this pattern can be seen to generalise to higher number of gravitons as well. This suggests that the solution to (\ref{rec_rel_A}) is given by
\be\label{sol_A}
A(\mathcal{K})=\sum_{\gamma\in\Gamma_{\mathcal{K}}}\prod_{i\in\mathcal{K}}\ket{i}{q}^{2(\alpha_i-2)}\prod_{\langle jk\rangle\in\gamma}\frac{\bra{j}{k}}{\ket{j}{k}},
\ee
where the first sum runs over paths in $\mathcal{K}$ of length $|\mathcal{K}|-1$ without repeting edges (trees), $\alpha_i$ is the number of edges at vertex $i$ and the second product is taken over vertices of the path. A combinatorial proof of the fact that (\ref{sol_A}) solves (\ref{rec_rel_A}) was already spelled out in e.g. \cite{Krasnov:2013wsa} and will not be repeated here.

\subsection{MHV amplitude formula}

It turns out that (\ref{sol_A}) contains all the essential features of the MHV graviton scattering amplitude formula. Consider the amplitude for two positive helicity gravitons with momentum spinors $x^A, y^A$ and $n$ negative helicity states, labelled as in the previous subsection. The MHV amplitude then turns out to be given by
\be\label{MHV}
{\cal M}^{MHV}(x,y,\{1,\ldots,n\}) = 2\im \left(\frac{\kappa}{2}\right)^n \ket{x}{y}^6 \sum_{\gamma\in\Gamma_{\mathcal{K}}}\prod_{i\in\mathcal{K}}\ket{i}{x}^{\alpha_i-2}\ket{i}{y}^{\alpha_i-2}\prod_{\langle jk\rangle\in\gamma}\frac{\bra{j}{k}}{\ket{j}{k}}.
\ee
Note that, apart from the pre factor of $\ket{x}{y}^6$ this is essentially the same formula as (\ref{sol_A}), except that it became generalised with the reference spinor $q$ being replaced in a symmetric fashion by two spinors $x,y$. The formula (\ref{MHV}) could be guessed from (\ref{sol_A}) by noting that in the limit $x\to y\to q$ when the two positive helicity gravitons become collinear and the the MHV amplitude develops a singularity, the sought formula should become proportional to the function $A(\{1,\ldots,n\})$ appearing in the all negative current. The sum over trees in (\ref{MHV}) is the most obvious function of $x,y$ with that property. The pre factor can then be easily determined from the behaviour under the little group, as well as dimensional considerations. The numerical factor can be determined by computing the 4-graviton case explicitly, which is not a difficult computation if one uses spinors. After the formula (\ref{MHV}) is guessed, it can be proven using the BCFW recursion relation, see \cite{Nguyen:2009jk} for such a proof. This sum over trees function of $x,y$ in (\ref{MHV}) first appeared in the already mentioned paper \cite{Bern:1998sv}, which also contains the formula (\ref{MHV}). 

Thus, to summarise the logic, once the all-negative current is computed using the Berends-Giele recursion, the MHV formula can be guessed and then checked using the BCFW recursion. This can be viewed as the simplest available derivation of the MHV formula (\ref{MHV}). 

\subsection{Analytic continuation and BCFW recursion}

To finish with our story for the all-negative current, we show that it satisfies an analog of the BCFW recursion relation. This is not surprising, because the MHV formula (\ref{MHV}) can be written in terms of an object that generalises the all-negative current, and the MHV amplitude satisfies BCFW, so the current must also satisfy it. Here we derive this BCWF-type recursion from Feynman diagrams. Aspects of what appears in this subsection was already spelled out in \cite{Krasnov:2013wsa}, but the presented here Feynman diagram derivation is new. 

The idea is to perform the BCFW-type shift on one of the negative on-shell legs. Thus, we shift the unprimed momentum spinor of the first graviton {\it in the direction of the reference spinor} 
$$1_z=1-zq,\qquad 1_z'=1'.$$ 
We do not need to shift any other momentum spinors because we work with a current, and so we have no intention to preserve the sum of all the momenta. It is only important that all the gravitons continue to be null. The particular form of the shift also guarantees that the helicity spinor for the shifted graviton is unchanged, so one can continue to use the already known expression for the off-shell current. Denoting the set of momentum spinors with $1$ shifted in this way by
$$\qquad\cK_z=\{1_z\}\sqcup\cK\backslash\{1\}$$
the function $A({\mathcal K})$ can easily be seen to depend on $z$ in the following way:
\be\label{A-z}
A(\cK_z)=\frac{1}{\ket{1}{q}^4}\prod_{i\in\cK\backslash\{1\}}\frac{1}{\ket{i}{q}^4}\sum_{T\in\cT(\cK)}\prod_{\stackrel{(jk)\in E(T)}{j,k\neq 1}}\frac{\bra{j}{k}}{\ket{j}{k}}\ket{j}{q}^2\ket{k}{q}^2\prod_{\langle 1l\rangle\in E(T)}\frac{\bra{1}{l}}{\ket{1}{l}-z\ket{q}{l}}\ket{1}{q}^2\ket{l}{q}^2.
\ee
We thus see that $A(\cK_z)$ has only simple poles at $z_l=\ket{1}{l}/\ket{q}{l}$. It is hard to compute the residues directly from this expression for $A(\cK_z)$. But we know that the current is obtained as a sum of Feynman diagrams, and these give a convenient way to obtain the residues. Indeed, the only diagrams with such poles are those in which the leg $1$  connects directly to $l$. This is because if the leg $1$ connects to a set $J$ of momenta already grouped together, the pole will be more complicated one of the form 
$$\sum_{j\in J} \ket{1}{j}\bra{1}{j} = z \sum_{j\in J} \ket{q}{j}\bra{1}{j}.$$
Since there are no such poles in the explicit expression for the current, we conclude that they must all cancel out, with only the poles present in (\ref{A-z}) remaining. We can then easily compute the residues at the poles $z_l=\ket{1}{l}/\ket{q}{l}$ from the corresponding Feynman diagrams. We get for the shifted current (suppressing all spinor indices)
$$J(\cK_z)=\sum_{l\in\cK\backslash\{1\}}\Big(\textnormal{sum over diagrams where }1_z\textnormal{ connects to }l\Big)+\cdots=\sum_{l\in\cK\backslash\{1\}}\frac{\bra{1}{l}\ket{l}{q}}{\ket{1}{q}^2}\frac{J(\cK^{\hat l}\backslash\{1\})}{z-z_l}+\cdots,$$
where $\hat l$ is defined by
$$\hat l\hat l'=1_{z_l}1'+ll'=l(l'+\frac{\ket{1}{q}}{\ket{l}{q}}1')$$
and $\cK^{\hat l}=\{\hat l\}\sqcup\cK\backslash\{l\}.$ This clearly identifies the poles of interest as well as their residues. 

We can now compute the current from its known residues, knowing also that the $z\to\infty$ behaviour of $A(\cK_z)$ and, therefore, of $J^-(\cK_z)$ is $1/z$. The residue theorem then gives us
\be\nonumber
\frac{1}{2\pi\im}\oint_{C_\infty}J(\cK_z)\frac{dz}{z}=-\Res(J(\cK_z)/z;0)-\sum_{z_l}\frac{1}{z_l}\Res(J(\cK_z);z_l)
\cr
=-J(\cK)+\sum_{l\in\cK\backslash\{1\}}\frac{\bra{1}{l}}{\ket{1}{l}}\frac{\ket{l}{q}^2}{\ket{1}{q}^2}J(\cK^{\hat l}\backslash\{1\})=0
\ee
$$\Longrightarrow\qquad A(\cK)=\sum_{l\in\cK\backslash\{1\}}\frac{\bra{1}{l}}{\ket{1}{l}}\frac{\ket{l}{q}^2}{\ket{1}{q}^2}A(\cK^{\hat l}\backslash\{1\}).$$
This is a BCFW-type recursion relation, but here for the Berends-Giele current. In \cite{Krasnov:2013wsa} we also gave a purely combinatorial proof of the validity of this recursion. But the presented here "physical" derivation shows the power of availability of access to Feynman diagrams. 

\subsection{Determinant representation}

Given that the all negative current is given by a sum over trees, a simple application of the matrix tree theorem allows to represent this current as a certain determinant. This has already been discussed in \cite{Krasnov:2013wsa}, so we just state the formula. 

Consider a complete graph whose vertices are elements from our set $\cK$. This is a graph where every vertex is connected to every other vertex (by exactly one edge). We now define a weighted Laplacian matrix for our complete graph, by associating 
\be
w_{ij} = \left\{ \lower2ex\vbox{ \hbox{$-\frac{\bra{i}{j}}{\ket{i}{j}}\ket{i}{q}^2\ket{j}{q}^2 ,\quad i\not=j.$}\hbox{ $\sum_{k\not=i} \frac{\bra{i}{k}}{\ket{i}{k}}\ket{i}{q}^2\ket{k}{q}^2, \quad i=j,$}} \right.
\ee
Then the determinant of the submatrix $W'$ of $W$ obtained by deleting $i$-th row and $i$-th column is independent of $i$ can can be denoted as $|W'|$. It is given by the following sum over trees
\be
|W'| = \sum_{T\in\cT(\cK)}\prod_{\langle ij\rangle\in E(T)} w_{ij}.
\ee
Looking at (\ref{sol_A}) we see that a multiple of our current is expressible as such a matrix determinant:
\be\label{A-det}
A(\cK) \prod_{i\in\cK}\ket{i}{q}^4 = |W'|.
\ee

\section{All minus one plus current recursion}
\label{sec:one-plus}

Here we consider a more non-trivial problem of writing a recursion relation for the all minus one plus current. This can then be composed with the final plus external graviton to yield the MHV amplitude. We take the auxiliary spinors of all negative helicity gravitons to be equal to the momentum spinor of the positive helicity graviton. We will denote the positive helicity graviton by $0$, and continue to number the negative helicity states from $1$ to $n$. To write our recursion relation it will be important to keep the positive helicity graviton slightly off the zero mass shell, see (\ref{massive}). 

We will continue to denote the reference spinor, which has been chosen to be the momentum of particle $0$, by $q^A$. Thus, the polarisation spinor for the positive helicity particle is given by
\be\label{pol-plus}
\epsilon^+(0)_{ABCD'} = \frac{1}{M} \frac{q_A q_B q_C p_{D'}}{\bra{p}{q}}.
\ee
The polarisation spinor is thus still proportional to $q_A q_B q_C$ as for all the negative helicity gravitons, which implies some simplifications. For example, it is easy to see that one can only glue this positive polarisation state with any of the negative polarisations by inserting them into the $\delta$-legs of the vertex. This means that a similar recursion relation to the one considered in the previous section will be possible. 

\subsection{Glueing the positive and a negative helicity graviton}

The easiest way to see how the recursion starts is by considering how our positive helicity graviton gets put together with some negative helicity state in the vertex. In contrast to the case of the all negative current, we found it convenient not to include the final propagator in the definition of the current. We thus rewrite the positive helicity spinor (\ref{pol-plus}) in the following suggestive way
\be\label{J0}
J(0)_{ABCD'} \equiv 2\im M^2 \epsilon^+(0)_{ABCA'} = -2\im M q_A q_B q_C 0_{DD'} q^D B(0),
\ee
where
$$B(0)=\frac{1}{M^2}.$$
The factor of $2\im M^2$ in (\ref{J0}) is the inverse propagator for the positive helicity graviton with $k_0^2=2M^2$. 

We now put together this zeroth order $B$-current together with a zeroth order $A$-current for a negative helicity graviton. The $B$-current has to be multiplied by the corresponding propagator first. The application of the $\delta$-operator to the $B$ current times the propagator gives
\be
\frac{1}{2\im M^2} (\delta J(0))^{ABA'B'} = \frac{1}{\im M} q^A q^B q^E q^F 0_E{}^{A'} 0_F{}^{B'} B(0).
\ee
We then apply the double commutator to this and (\ref{dJ1}) to get
\be
\frac{1}{2\im M^2}  [\![ \delta J(0), \delta J(1) ]\!]^{ABCD} =  q^A q^B q^C q^D  (q^E q^F 0_E{}^{A'} 0_F{}^{B'})( q^P q^Q 1_{PA'} 1_{QB'}) B(0) A(1).
\ee
We finally apply to this the final derivative that gives a factor of $\im(0+1)_{AA'}$. There is no need to add the propagator at the end with our definition for this current. Thus we get
\be
J(0,1)_{ABCD'} = -2\im M \left(\frac{\kappa}{2}\right) q_A q_B q_C (0+1)_{DD'} q^D B(0,1),
\ee
where
\be\label{B01}
B(0,1) = \frac{M^{2}}{\bra{p}{q}^2} \left( \ket{1}{q} \bra{1}{p} \right)^2 B(0) A(1) = \frac{\bra{1}{p}^2}{\bra{p}{q}^2 \ket{1}{q}^2}.
\ee  
Thus, unlike $B(0)$, this current is independent of $M$. 

\subsection{General recursion}

The general recursion for the $B$-current follows the established above pattern. Thus, we can write it as
\be
J(0,\cK)_{ABCD'} = -2\im M \left(\frac{\kappa}{2}\right)^{|\cK|} q_A q_B q_C (0+\sum_{\cK} k)_{DD'} q^D B(0,\cK).
\ee
Note that the zeroth momentum in the sum above can be dropped as compared to all others, because it contributes a term of the order $M^2$, to be neglected in the $M\to 0$ limit. The $B$-coefficients satisfy the following recursion relation:
\be\nonumber
B(0,\mathcal{K})=\sum_{\mathcal{I}\sqcup\mathcal{J}=\mathcal{K}}\frac{B(0,\mathcal{I})A(\mathcal{J})}{\sum_{i\in\mathcal{I}}(iq)[iq]+\sum_{i<j\in\mathcal{I}}\ket{i}{j}\bra{i}{j}}\left(\sum_{a\in\mathcal{I},b\in\mathcal{J}}\ket{a}{q}\ket{b}{q}\bra{a}{b}\right)^2+\frac{A(\mathcal{K})}{\bra{p}{q}^2}\left(\sum_{a\in\mathcal{K}}\ket{a}{q}\bra{a}{p}\right)^2
\ee
The first set of terms here comes from glueing a current $J(0,\cI)$ with non-empty set $\cI$ to an $A$-current for some set $\cJ$. It is precisely the recursion of the type that we already analysed before for the $A$-current, except that now the propagator for the $B$-current has to be added before this current is glued to the vertex, not after. The last term here comes from the possibility of glueing the $A$-current for the whole set $\cK$ directly to the positive helicity graviton. It follows the pattern seen in (\ref{B01}). 
 
The first few examples of application of the above recursion are
 $$B(0,1)=A(1)\frac{\ket{1}{q}\bra{1}{p}^2}{\bra{1}{q}}\frac{\ket{1}{q}\bra{1}{q}}{\bra{p}{q}^2},$$
 \be \nonumber
 B(0,1,2)=A(1,2)\left(\frac{\ket{1}{q}\bra{1}{p}^2}{\bra{1}{q}}+\frac{\ket{2}{q}\bra{2}{p}^2}{\bra{2}{q}}\right)\frac{\ket{1}{q}\bra{1}{q}+\ket{2}{q}\bra{2}{q}+\ket{1}{2}\bra{1}{2}}{\bra{p}{q}^2}-A(1,2)\bra{1}{2}^2 \frac{\ket{1}{q}\ket{2}{q}}{\bra{1}{q}\bra{2}{q}}.
 \ee
We wrote both in a suggestive form. 

One can now show that the general $B$-current has two terms. One term is proportional to the sum of all momenta squared, and includes the factor of $1/[pq]^2$. This term will vanish when composed with the final positive helicity state (because of the momenta conservation). The other term does not depend on the auxiliary spinor $p$
\be
B(0,\mathcal{K})=A(\mathcal{K})C(0,\mathcal{K})\frac{\sum_{i\in\mathcal{K}}\ket{i}{q}\bra{i}{q}+\sum_{i<j\in\mathcal{K}}\ket{i}{j}\bra{i}{j}}{\bra{p}{q}^2}+S(0,\mathcal{K}).
\ee
Substituting this into the recursion relation for the $B$-current and then using the recursion relation $A$-current it is possible to show (see Appendix) that 
\be\label{C}
C(0,\mathcal{K})=\sum_{i\in\mathcal{K}}\frac{\ket{i}{q}\bra{i}{p}^2}{\bra{i}{q}}
\ee
and that the following recursion relation for the $S$-amplitudes holds:
\be\label{recursion-S}
S(0,{\cal K}) = \sum_{{\cal I}\cup{\cal J}={\cal K}} \frac{S(0,{\cal I})}{\sum_{i\in {\cal I}} \ket{i}{q}\bra{i}{q} + \sum_{i<j\in{\cal I}} \ket{i}{j}\bra{i}{j}}  A({\cal J}) \left( \sum_{i\in{\cal I}} \sum_{j\in {\cal J}} \ket{i}{q}\ket{j}{q} \bra{i}{j} \right)^2 \\ \nonumber
- A({\cal K}) \sum_{i<j \in{\cal K}} \frac{\ket{i}{q}\ket{j}{q}}{\bra{i}{q}\bra{j}{q}} \bra{i}{j}^2.
\ee
This recursion should be started with $S(0)=S(0,1)=0$. The first non-trivial amplitude is then for $n=2$ and arises directly from the last term in (\ref{recursion-S}). Thus, we have:
\be\label{S-2}
S(0,1,2)=-A(1,2) \frac{\ket{1}{q}\ket{2}{q}}{\bra{1}{q}\bra{2}{q}} \bra{1}{2}^2.
\ee
It is not hard to see that composing the resulting current with the final positive helicity state gives the correct $--++$ graviton amplitude. We have also checked that the 5-graviton amplitude that can be obtained from (\ref{recursion-S}) is correct. Unfortunately, we were unable to solve this recursion in general. 

\section{Discussion}

There are two main results of this paper. The first one is the derivation of the recursion relation (\ref{rec_rel_A}) from Feynman rules of the gauge-theoretic formulation of gravity. The same recursion relation was also arrived at in \cite{Bern:1998sv}, but we believe that our Feynman rules lead to this relation in a more transparent and efficient way. The solution (\ref{sol_A}) to (\ref{rec_rel_A}) already appeared in \cite{Bern:1998sv} and in this sense it is not a new result. However, we think the fact that the Feynman rule derived recursion relation directly leads to a sum over trees representation of the graviton amplitudes cannot be overemphasised. It appears to not be universally appreciated by the amplitude community, and we hope this paper will help to rectify this. 

Our other main result is the derivation of the new recursion (\ref{recursion-S}) for the current with all but one negative helicity gravitons. This is the current that can be directly used for computing the MHV graviton amplitudes, as it does not vanish when the last leg is set on-shell. We have obtained a recursion relation for the main building block of this current, which is independent of any unphysical reference spinors, or factors of $M$. The new recursion relation differs from (\ref{rec_rel_A}) in the appearance of the inhomogeneous term. It seems that it is this term that makes life complicated, and prevents from spotting any simple pattern for this current. It would be very interesting to find a closed-form solution of (\ref{recursion-S}), because it contains more information than the known purely on-shell MHV amplitudes. 

\section*{Acknowledgements} The authors were supported by an ERC Starting Grant 277570-DIGT.

\section*{Appendix: Proof of the $S$-recursion}

Here we give a proof of the recursion (\ref{recursion-S}) together with the expression (\ref{C}) for $C(0,\cK)$ using the recursions for the $B$ and $A$-currents. Let's assume, by induction, that $C(0,\mathcal{K})$ is given by
$$C(0,\mathcal{K})=\sum_{i\in\mathcal{K}}\frac{\ket{i}{q}\bra{i}{p}^2}{\bra{i}{q}}$$
for $|\mathcal{K}|<n$. Then, for $|\mathcal{K}|=n$ we get
\be
S(0,\mathcal{K})+A(\mathcal{K})C(0,\mathcal{K})\frac{\sum_{i\in\mathcal{K}}\ket{i}{q}\bra{i}{q}+\sum_{i<j\in\mathcal{K}}\ket{i}{j}\bra{i}{j}}{\bra{p}{q}^2}
 \\ \nonumber
 =\sum_{\mathcal{I}\sqcup\mathcal{J}=\mathcal{K}}\frac{B(0,\mathcal{I})A(\mathcal{J})}{\sum_{i\in\mathcal{I}}\ket{i}{q}\bra{i}{q}+\sum_{i<j\in\mathcal{I}}\ket{i}{j}\bra{i}{j}}\left(\sum_{a\in\mathcal{I},b\in\mathcal{J}}\ket{a}{q}\ket{b}{q}\bra{a}{b}\right)^2+\frac{A(\mathcal{K})}{\bra{p}{q}^2}\left(\sum_{i\in\mathcal{K}}\ket{i}{q}\bra{i}{p}\right)^2
\\ \nonumber
=\sum_{\mathcal{I}\sqcup\mathcal{J}=\mathcal{K}}\frac{S(0,\mathcal{I})A(\mathcal{J})}{\sum_{i\in\mathcal{I}}\ket{i}{q}\bra{i}{q}+\sum_{i<j\in\mathcal{I}}\ket{i}{j}\bra{i}{j}}\left(\sum_{a\in\mathcal{I},b\in\mathcal{J}}\ket{a}{q}\ket{b}{q}\bra{a}{b}\right)^2+\frac{A(\mathcal{K})}{\bra{p}{q}^2}\left(\sum_{i\in\mathcal{K}}\ket{i}{q}\bra{i}{p}\right)^2
\\ \nonumber
+\sum_{\mathcal{I}\sqcup\mathcal{J}=\mathcal{K}}A(\mathcal{I})A(\mathcal{J})\left(\sum_{a\in\mathcal{I},b\in\mathcal{J}}\ket{a}{q}\ket{b}{q}\bra{a}{b}\right)^2\frac{1}{\bra{p}{q}^2}\sum_{i\in\mathcal{I}}\frac{\ket{i}{q}\bra{i}{p}^2}{\bra{i}{q}}.
\ee
Now,
\be
\sum_{\mathcal{I}\sqcup\mathcal{J}=\mathcal{K}}A(\mathcal{I})A(\mathcal{J})\left(\sum_{a\in\mathcal{I},b\in\mathcal{J}}\ket{a}{q}\ket{b}{q}\bra{a}{b}\right)^2\frac{1}{\bra{p}{q}^2}\sum_{i\in\mathcal{I}}\frac{\ket{i}{q}\bra{i}{p}^2}{\bra{i}{q}}
\\ \nonumber
=\sum_{\mathcal{I}\sqcup\mathcal{J}=\mathcal{K},|\mathcal{I}|\leq|\mathcal{J}|}A(\mathcal{I})A(\mathcal{J})\left(\sum_{a\in\mathcal{I},b\in\mathcal{J}}\ket{a}{q}\ket{b}{q}\bra{a}{b}\right)^2\frac{1}{\bra{p}{q}^2}\left[\sum_{i\in\mathcal{I}}\frac{\ket{i}{q}\bra{i}{p}^2}{\bra{i}{q}}+\sum_{i\in\mathcal{J}}\frac{\ket{i}{q}\bra{i}{p}^2}{\bra{i}{q}}\right]
\\ \nonumber
=A(\mathcal{K})\sum_{k\in\mathcal{K}}\frac{\ket{k}{q}\bra{k}{p}^2}{\bra{k}{q}}\frac{\sum_{i\in\mathcal{K}}\ket{i}{q}\bra{i}{q}+\sum_{i<j\in\mathcal{K}}\ket{i}{j}\bra{i}{j}}{\bra{p}{q}^2}-A(\mathcal{K})\frac{\sum_{i\in\mathcal{K}}\ket{i}{q}\bra{i}{q}}{\bra{p}{q}^2}\sum_{k\in\mathcal{K}}\frac{\ket{k}{q}\bra{k}{p}^2}{\bra{k}{q}}
\ee
and
\be
\sum_{i\in\mathcal{K}}\ket{i}{q}\bra{i}{q}\sum_{k\in\mathcal{K}}\frac{\ket{k}{q}\bra{k}{p}^2}{\bra{k}{q}}-\left(\sum_{i\in\mathcal{K}}\ket{i}{q}\bra{i}{p}\right)^2=\sum_{i,j\in\mathcal{K}}\left(\ket{i}{q}\bra{i}{q}\frac{\ket{j}{q}\bra{j}{p}^2}{\bra{j}{q}}-\ket{i}{q}\bra{i}{p}\ket{j}{q}\bra{j}{p}\right)
\\ \nonumber
=-\bra{p}{q}\sum_{i,j\in\mathcal{K}}\frac{\ket{i}{q}\ket{j}{q}\bra{j}{p}}{\bra{j}{q}}\bra{i}{j}=\bra{p}{q}^2\sum_{i<j\in\mathcal{K}}\frac{\ket{i}{q}\ket{j}{q}}{\bra{i}{q}\bra{j}{q}}\bra{i}{j}^2.
\ee
Using these identities we get
 \be\nonumber
 S(0,\mathcal{K})-A(\mathcal{K})C(0,\mathcal{K})\frac{\sum_{i\in\mathcal{K}}\ket{i}{q}\bra{i}{q}+\sum_{i<j\in\mathcal{K}}\ket{i}{j}\bra{i}{j}}{\bra{p}{q}^2}
  \\ \nonumber
 =\sum_{\mathcal{I}\sqcup\mathcal{J}=\mathcal{K}}\frac{S(0,\mathcal{I})A(\mathcal{J})}{\sum_{i\in\mathcal{I}}\ket{i}{q}\bra{i}{q}+\sum_{i<j\in\mathcal{I}}\ket{i}{j}\bra{i}{j}}\left(\sum_{a\in\mathcal{I},b\in\mathcal{J}}\ket{a}{q}\ket{b}{q}\bra{a}{b}\right)^2-\frac{A(\mathcal{K})}{\bra{p}{q}^2}\left(\sum_{i\in\mathcal{K}}\ket{i}{q}\bra{i}{p}\right)^2
 \\ \nonumber 
 -A(\mathcal{K})\sum_{k\in\mathcal{K}}\frac{\ket{k}{q}\bra{k}{p}^2}{\bra{k}{q}}\frac{\sum_{i\in\mathcal{K}}\ket{i}{q}\bra{i}{q}+\sum_{i<j\in\mathcal{K}}\ket{i}{j}\bra{i}{j}}{\bra{p}{q}^2}+A(\mathcal{K})\frac{\sum_{i\in\mathcal{K}}\ket{i}{q}\bra{i}{q}}{\bra{p}{q}^2}\sum_{k\in\mathcal{K}}\frac{\ket{k}{q}\bra{k}{p}^2}{\bra{k}{q}}
 \\ \nonumber
 =\sum_{\mathcal{I}\sqcup\mathcal{J}=\mathcal{K}}\frac{S(0,\mathcal{I})A(\mathcal{J})}{\sum_{i\in\mathcal{I}}\ket{i}{q}\bra{i}{q}+\sum_{i<j\in\mathcal{I}}\ket{i}{j}\bra{i}{j}}\left(\sum_{a\in\mathcal{I},b\in\mathcal{J}}\ket{a}{q}\ket{b}{q}\bra{a}{b}\right)^2
 \\ \nonumber
 -A(\mathcal{K})\sum_{k\in\mathcal{K}}\frac{\ket{k}{q}\bra{k}{p}^2}{\bra{k}{q}}\frac{\sum_{i\in\mathcal{K}}\ket{i}{q}\bra{i}{q}+\sum_{i<j\in\mathcal{K}}\ket{i}{j}\bra{i}{j}}{\bra{p}{q}^2}
 \\ \nonumber
 +A(\mathcal{K})\frac{1}{\bra{p}{q}^2}\sum_{i,j\in\mathcal{K}}\left(\ket{i}{q}\bra{i}{q}\frac{\ket{j}{q}\bra{j}{p}^2}{\bra{j}{q}}-\ket{i}{q}\bra{i}{p}\ket{j}{q}\bra{j}{p}\right).
\ee
Finally, combining the $A(\cK)$ terms here we get 
\be\nonumber
 =\sum_{\mathcal{I}\sqcup\mathcal{J}=\mathcal{K}}\frac{S(0,\mathcal{I})A(\mathcal{J})}{\sum_{i\in\mathcal{I}}\ket{i}{q}\bra{i}{q}+\sum_{i<j\in\mathcal{I}}\ket{i}{j}\bra{i}{j}}\left(\sum_{a\in\mathcal{I},b\in\mathcal{J}}\ket{a}{q}\ket{b}{q}\bra{a}{b}\right)^2
 -A(\mathcal{K})\frac{1}{\bra{p}{q}}\sum_{i,j\in\mathcal{K}}\frac{\ket{i}{q}\ket{j}{q}\bra{j}{p}}{\bra{j}{q}}\bra{i}{j}
 \\ \nonumber
 -A(\mathcal{K})\sum_{k\in\mathcal{K}}\frac{\ket{k}{q}\bra{k}{p}^2}{\bra{k}{q}}\frac{\sum_{i\in\mathcal{K}}\ket{i}{q}\bra{i}{q}+\sum_{i<j\in\mathcal{K}}\ket{i}{j}\bra{i}{j}}{\bra{p}{q}^2}
 \\ \nonumber
 =\sum_{\mathcal{I}\sqcup\mathcal{J}=\mathcal{K}}\frac{S(0,\mathcal{I})A(\mathcal{J})}{\sum_{i\in\mathcal{I}}\ket{i}{q}\bra{i}{q}+\sum_{i<j\in\mathcal{I}}\ket{i}{j}\bra{i}{j}}\left(\sum_{a\in\mathcal{I},b\in\mathcal{J}}\ket{a}{q}\ket{b}{q}\bra{a}{b}\right)^2
 +A(\mathcal{K})\sum_{i<j\in\mathcal{K}}\frac{\ket{i}{q}\ket{j}{q}}{\bra{i}{q}\bra{j}{q}}\bra{i}{j}^2
 \\ \nonumber
 -A(\mathcal{K})\sum_{k\in\mathcal{K}}\frac{\ket{k}{q}\bra{k}{p}^2}{\bra{k}{q}}\frac{\sum_{i\in\mathcal{K}}\ket{i}{q}\bra{i}{q}+\sum_{i<j\in\mathcal{K}}\ket{i}{j}\bra{i}{j}}{\bra{p}{q}^2}
 \ee
Therefore we get the expected formula (\ref{C}) for $C(0,\mathcal{K})$ and a remaining recursion relation (\ref{recursion-S}) for $S(0,\mathcal{K})$.


\begin{thebibliography}{99}

\bibitem{Krasnov:2011pp} 
  K.~Krasnov,
  ``Pure Connection Action Principle for General Relativity,''
  Phys.\ Rev.\ Lett.\  {\bf 106}, 251103 (2011)
  [arXiv:1103.4498 [gr-qc]].

\bibitem{Krasnov:2011up} 
  K.~Krasnov,
  ``Gravity as a diffeomorphism invariant gauge theory,''
  Phys.\ Rev.\ D {\bf 84}, 024034 (2011)
  [arXiv:1101.4788 [hep-th]].
  
\bibitem{Delfino:2012zy} 
  G.~Delfino, K.~Krasnov and C.~Scarinci,
  ``Pure Connection Formalism for Gravity: Linearized Theory,''
  arXiv:1205.7045 [hep-th].
  
\bibitem{Delfino:2012aj} 
  G.~Delfino, K.~Krasnov and C.~Scarinci,
  ``Pure connection formalism for gravity: Feynman rules and the graviton-graviton scattering,''
  arXiv:1210.6215 [hep-th].
  
\bibitem{Berends:1987me}
  F.~A.~Berends and W.~T.~Giele,
  ``Recursive Calculations for Processes with n Gluons,''
  Nucl.\ Phys.\  B {\bf 306}, 759 (1988).
 
\bibitem{Parke:1986gb}
  S.~J.~Parke and T.~R.~Taylor,
  ``An Amplitude for $n$ Gluon Scattering,''
  Phys.\ Rev.\ Lett.\  {\bf 56}, 2459 (1986).
 
\bibitem{Bern:1998sv}
  Z.~Bern, L.~J.~Dixon, M.~Perelstein and J.~S.~Rozowsky,
  ``Multi-leg one-loop gravity amplitudes from gauge theory,''
  Nucl.\ Phys.\  B {\bf 546}, 423 (1999)
  [arXiv:hep-th/9811140].
  
\bibitem{Nguyen:2009jk} 
  D.~Nguyen, M.~Spradlin, A.~Volovich and C.~Wen,
  ``The Tree Formula for MHV Graviton Amplitudes,''
  JHEP {\bf 1007}, 045 (2010)
  [arXiv:0907.2276 [hep-th]].
  
\bibitem{Hodges:2012ym} 
  A.~Hodges,
  ``A simple formula for gravitational MHV amplitudes,''
  arXiv:1204.1930 [hep-th].
  
\bibitem{Feng:2012sy} 
  B.~Feng and S.~He,
  ``Graphs, determinants and gravity amplitudes,''
  JHEP {\bf 1210}, 121 (2012)
  [arXiv:1207.3220 [hep-th]].
  
\bibitem{Krasnov:2013wsa} 
  K.~Krasnov and C.~Scarinci,
  ``Weighted Laplacians, cocycles and recursion relations,''
  JHEP {\bf 1311}, 040 (2013)
  [arXiv:1310.0653 [hep-th]].
  
\bibitem{Britto:2005fq} 
  R.~Britto, F.~Cachazo, B.~Feng and E.~Witten,
  ``Direct proof of tree-level recursion relation in Yang-Mills theory,''
  Phys.\ Rev.\ Lett.\  {\bf 94}, 181602 (2005)
  [hep-th/0501052].
 
 
 
 
\bibitem{Berends:1988zp}
  F.~A.~Berends, W.~T.~Giele and H.~Kuijf,
  ``On relations between multi - gluon and multigraviton scattering,''
  Phys.\ Lett.\  B {\bf 211}, 91 (1988).
  
\bibitem{Kawai:1985xq}
  H.~Kawai, D.~C.~Lewellen and S.~H.~H.~Tye,
  ``A Relation Between Tree Amplitudes of Closed and Open Strings,''
  Nucl.\ Phys.\  B {\bf 269}, 1 (1986).
  
\bibitem{Bedford:2005yy}
  J.~Bedford, A.~Brandhuber, B.~J.~Spence and G.~Travaglini,
  ``A Recursion relation for gravity amplitudes,''
  Nucl.\ Phys.\  B {\bf 721}, 98 (2005)
  [arXiv:hep-th/0502146].
  
\bibitem{Cachazo:2005ca}
  F.~Cachazo and P.~Svrcek,
  ``Tree level recursion relations in general relativity,''
  arXiv:hep-th/0502160.
  
\bibitem{Mason:2008jy}
  L.~J.~Mason and D.~Skinner,
  ``Gravity, Twistors and the MHV Formalism,''
  Commun.\ Math.\ Phys.\  {\bf 294}, 827 (2010)
  [arXiv:0808.3907 [hep-th]].

  

    
  
  
 
  
  

\end{thebibliography}
\end{document}